\documentclass[prl,twocolumn,showpacs,floatfix,amsmath,amssymb,superscriptaddress] {revtex4-1}
\pdfoutput=1
\usepackage{latexsym}
\usepackage{graphicx}
\usepackage{times}
\usepackage{amsmath}
\usepackage{dcolumn}
\usepackage{latexsym,amsmath,amssymb,bm,euscript}
\newcommand{\beq}{\begin{equation}}
\newcommand{\eeq}{\end{equation}}
\newcommand{\bea}{\begin{eqnarray}}
\newcommand{\eea}{\end{eqnarray}}
\newcommand{\be}{\begin{equation}}
\newcommand{\ee}{\end{equation}}

\def\beq{\begin{equation}}
\def\eeq{\end{equation}}
\def\bea{\begin{eqnarray}}
\def\eea{\end{eqnarray}}

\begin{document}

\title{Phases of triangular lattice antiferromagnet near saturation}

\author{Oleg A. Starykh}
\affiliation{Department of Physics and Astronomy, University of Utah, Salt Lake
City, UT 84112}
\author{Wen Jin}
\affiliation{Department of Physics and Astronomy, University of Utah, Salt Lake
City, UT 84112}
\author{Andrey V. Chubukov}
\affiliation{Department of Physics, University of Wisconsin, Madison,
WI 53706}

\date{\today}

\begin{abstract}
We consider
 2D Heisenberg antiferromagnets on a triangular lattice with spatially anisotropic  interactions
 in a high magnetic field close to the saturation.
We show that this system  possess rich phase diagram in field/anisotropy plane
 due to competition between classical and quantum orders:
  an incommensurate non-coplanar spiral state, which is favored classically, and a commensurate co-planar state,
 which is stabilized by quantum fluctuations.
  We show that the transformation between these two states
   is highly non-trivial and involves  two intermediate  phases -- the phase with co-planar incommensurate spin order and the one
   with non-coplanar double-${\bf Q}$ spiral order.
 The transition between the two co-planar states is of commensurate-incommensurate type, not accompanied by softening of spin-wave excitations.
 We show that a different sequence of transitions holds in triangular antiferromagnets with exchange anisotropy, such as Ba$_3$CoSb$_2$O$_9$.
 \end{abstract}

\maketitle


{\bf Introduction.} The field of frustrated quantum magnetism witnessed a remarkable revival of interest in the last few years due to
rapid progress in synthesis of new materials and
 in understanding
 previously unknown states of matter.
 The two main lines of research
 in the field are searches
  for spin-liquid phases and
  for new ordered phases with highly non-trivial spin structures \cite{LB-review}.
 For the latter, the most promising system
  is a 2D Heisenberg antiferromagnet  on a triangular lattice in a finite magnetic field, as this system
   is known to possess an "accidental"  classical degeneracy: every classical spin configuration with a
  triad of neighboring spins satisfying
 ${\bf S}_{\bf r} + {\bf S}_{{\bf r} + {\bm \delta}_1} + {\bf S}_{{\bf r} + {\bm \delta}_2} = {\bf h}/(3 J)$, where $J$ is the exchange interaction, belongs to the
 ground state manifold.

\begin{figure}[h]
\begin{center}
\includegraphics[scale=0.25]{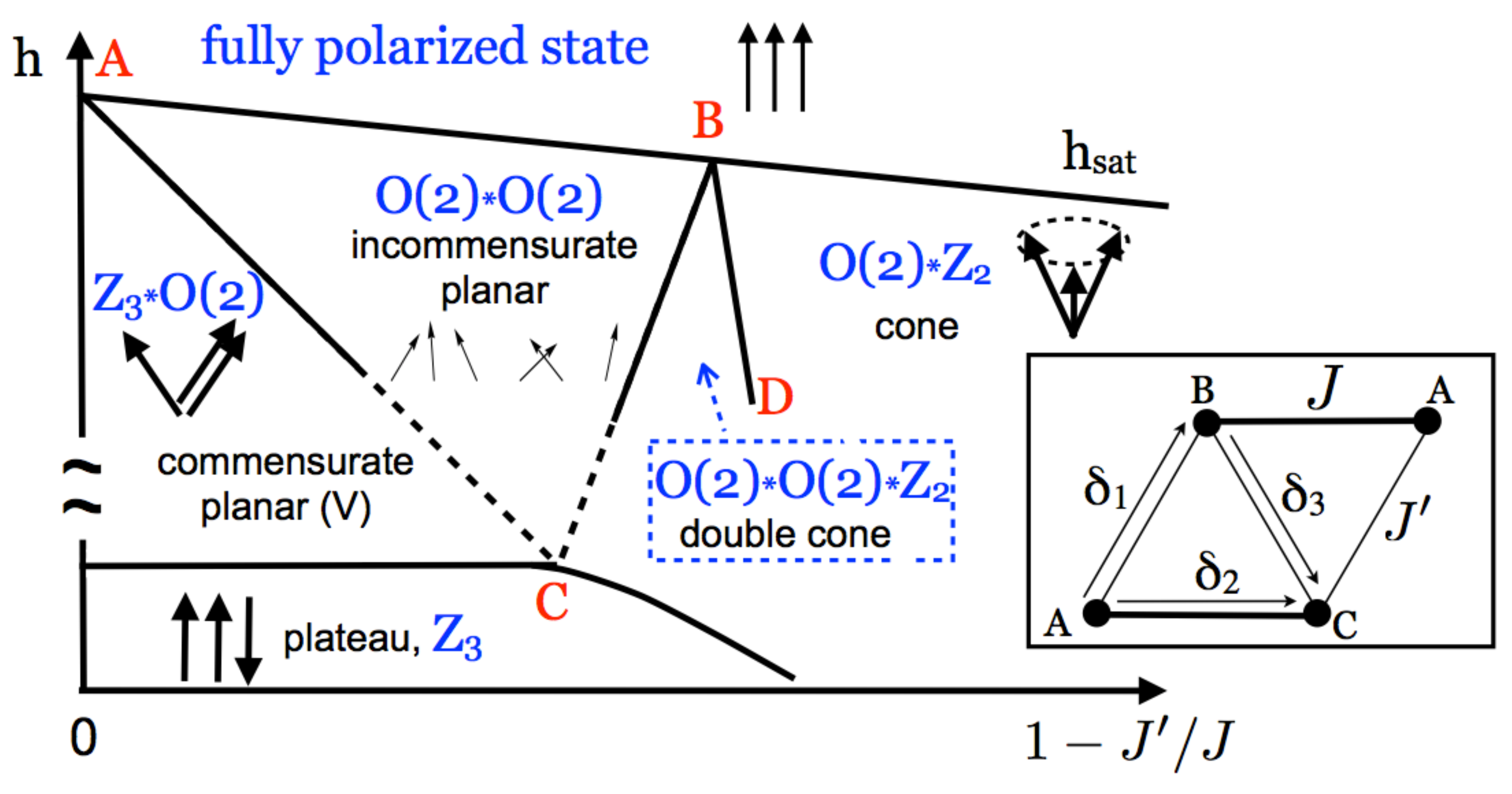}
\caption{
Phase diagram of the spatially anisotropic triangular lattice antiferromagnet with large $S$ near saturation field, as a function
 of spatial anisotropy of the interactions.  The phases at small and large anisotropy are commensurate co-planar  V-phase,
  which breaks $Z_3 \times O(2)$ symmetry, and incommensurate non-coplanar chiral cone phase, which breaks $Z_2 \times O(2)$ symmetry.
  In between, there are two incommensurate phases: a co-planar phase, which breaks $O(2) \times O(2)$ symmetry, and a
  non-coplanar double cone phase, which breaks $Z_2 \times O(2) \times O(2)$ symmetry.
  Line AC denotes the CI transition from the V phase to the incommensurate planar phase.
 The insert shows the geometry of the lattice exchange constant is $J$ on horizontal bonds (bold) and $J'$ on  diagonal  bonds (thin).}
  \label{fig:anis}
\end{center}
\end{figure}

 An infinite degeneracy, however, holds
 only for an ideal Heisenberg system with
 isotropic nearest-neighbor interaction.
  Real systems have either spatial anisotropy of exchange interactions, as in Cs$_2$CuCl$_4$ \cite{coldea2002,tokiwa} and
  Cs$_2$CuBr$_4$ \cite{ono2005,takano,zvyagin2014} for which
 the interaction $J$ on horizontal bonds is larger than $J'$ on diagonal bonds (see insert in Fig. \ref{fig:anis}),
  or exchange anisotropy in spin space,
 as in Ba$_3$CoSb$_2$O$_9$, for which $J_z < J_{\perp} = J$ (an easy plane anisotropy) \cite{shirata,tanaka,koutroulakis}.
 An anisotropy of either type breaks accidental degeneracy
   already at a classical level and for fields ${\bf h} = h \hat{z}$ slightly below the saturation field $h_{\rm sat}$
   selects a non-coplanar cone state
  with
 \begin{equation}
\langle {\bf S}_{\bf r} \rangle = (S -\rho) \hat{z} + \sqrt{2 S \rho} ( \cos[{\bf Q} \cdot {\bf r} + \varphi] \hat{x} + \sin[{\bf Q} \cdot {\bf r} + \varphi] \hat{y}),
\label{eq:umbrella}
\end{equation}
where $\rho \sim S (h_{\rm sat} - h)/h_{\rm sat}$ is the density of magnons (the {\em condensate fraction})
which determines the magnetization $M = S  - \rho$,
 $\varphi \in (0,2\pi)$  is a phase of a condensate,
and ${\bf Q} = (Q,0)$ is the ordering wave vector.
It is incommensurate
 with $Q= Q_i = 2 \cos^{-1}(-J'/2J)$ in the spatially anisotropic case $J' \neq J$ and commensurate
 with $Q= Q_0 = 4\pi/3$ for the easy-plane anisotropy
(in the last case, the values of ${\bf Q}_0 \cdot {\bf r} = 2\pi \nu/3 ~({\text{mod}} ~2\pi)$, with $\nu = \pm 1,0$).

 Quantum fluctuations are also known to lift accidental degeneracy,
  and do so  already in the {\em isotropic} system.  However,
  they select
    different ordered state, which is the co-planar, commensurate state with two parallel spins in every triad,
   often called the V state (Fig. \ref{fig:anis}) \cite{golosov,nikuni,griset}.

  This order is described by
  \begin{eqnarray}
\langle {\bf S}_{\bf r} \rangle &=& (S - 2\rho \cos^2[{\bf Q} \cdot {\bf r} + \theta]) \hat{z}
 + \sqrt{4 S \rho} \cos[{\bf Q} \cdot {\bf r} + \theta] \nonumber\\
 && \times \left(\cos {\varphi} \hat{x} + \sin{\varphi} \hat{y}\right),
\label{eq:V}
\end{eqnarray}
 where
 ${\bf Q} = {\bf Q}_0$,  $\rho = \rho_{{\bf Q}_0} + \rho_{-{\bf Q}_0}$ is
   the sum of
    two
    equal contributions
 from condensates with wave vectors $\pm {\bf Q}_0 = (\pm Q_0,0)$, $\varphi$ is
   a common phase of the two condensates,
 and $\theta$ is
   their
   {\em relative} phase. The values of $\theta$
   in the
   commensurate $V$ phase
   are
   constrained to
   $\theta = \pi \ell/3$,
  where
  $\ell = 0, 1, 2$ describe three distinct degenerate spin configurations
   (three choices to select two parallel spins in any triad, see Fig. \ref{fig:anis}).

   The issue we consider in this paper is how
  the system evolves
   at $h \leq h_{\rm sat}$
   from the co-planar $V$ state,
   selected by quantum fluctuations, to the
   non-coplanar cone state, selected by classical
  fluctuations,
  as the anisotropy
   increases.
 We show that this evolution is highly non-trivial and involves commensurate-incommensurate transition (CIT) and,
 in the case of $J-J'$ model, an intermediate double cone phase.

{\bf The phase diagrams.}
To begin, it is instructive to compare order parameter manifolds in the two phases.
   The
    order parameter manifold  in the V phase is $O(2) \times Z_3$  and that in the cone phase is $O(2) \times Z_2$.
    In both phases, a continuous $O(2)$ reflects a choice of the phase $\varphi$.
$Z_3$ in the V phase corresponds to choosing one of three values of $\theta$ in  (\ref{eq:V}), and
$Z_2$ in the cone phase is a chiral symmetry between left- and right-handed spiral orders (chiralities),
i.e. orders with $+Q$ and $-Q$ in (\ref{eq:umbrella}).
The symmetry breaking patterns in the two phases are not compatible, hence one should expect either  first-order transition(s) or
  an intermediate phase(s).  We show  that in $J-J'$ model the evolution occurs via two intermediate phases, see Fig.~\ref{fig:anis}.
 As $\delta J = J-J'$ increases, the V phase first undergoes a CIT
 at $\delta J_{c1} \sim (J /\sqrt{S})
 (h_{\rm sat} - h)/h_{\rm sat}$
 (line AC in Fig.~\ref{fig:anis}).
  The new phase remains co-planar, like in (\ref{eq:V}), but the phase $\theta$ becomes incommensurate and coordinate-dependent.
 and order parameter manifold extends to $O(2) \times O(2)$ (spontaneous selection of $\varphi$ and the origin of coordinates).
The incommensurate co-planar state
 exists up to a second critical
 $\delta J_{c2} \sim J/\sqrt{S}$, where the system breaks the $Z_2$ symmetry between the two condensates (line BC in Fig.~\ref{fig:anis})..
 At larger $\delta J$  the two condensates still develop,
one of them shifts to a new wave vector ${\bf {\bar Q}}$ and
 its magnitude gets smaller.
 The resulting state is a non-coplanar double cone state
 with order parameter manifold $O(2) \times O(2) \times Z_2$.
 Finally, at the third  critical anisotropy
 $\delta J_{c3} =  \delta J_{c2}[1 + O((h_{\rm sat} - h)/h_{\rm sat})]$
 the magnitude of the condensate at ${\bf {\bar Q}}$ vanishes and the double cone transforms into a single cone (line BD in Fig.~\ref{fig:anis}).
 \begin{figure}[h]
\begin{center}
\includegraphics[scale=0.25]{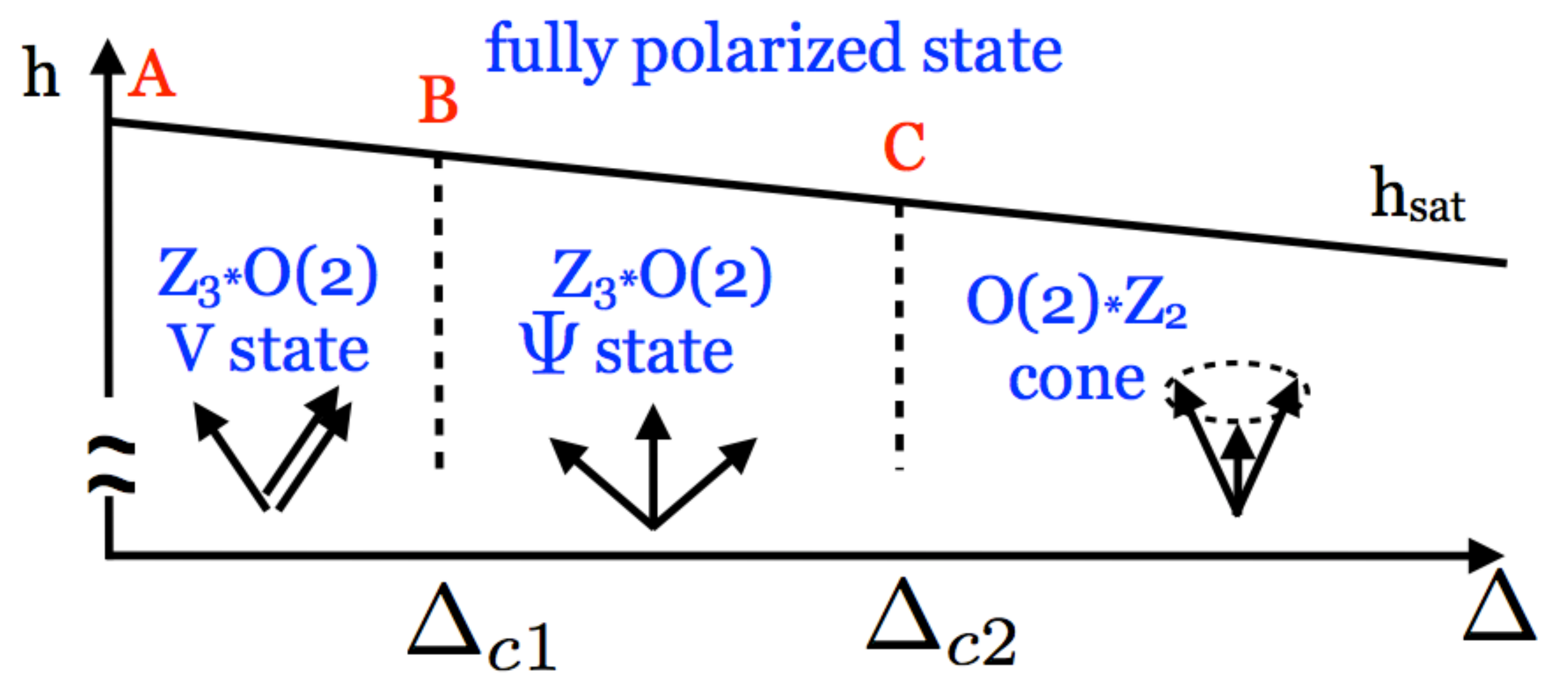}
\caption{The
phase diagram of the XXZ model in a magnetic field near a saturation value, $\Delta = (J - J_z)/J$.
 The cone and V states are the same as in Fig. \ref{fig:anis}, but the transformation from one phase to the other with increasing spin exchange anisotropy
  proceeds differently from the case of spatial exchange anisotropy and involves one intermediate co-planar commensurate phase
  with $\Psi$-like spin pattern.}
\label{fig:xxz}
\end{center}
\end{figure}

In systems  with easy-plane anisotropy $\Delta = (J - J_z)/J > 0$, the
 the ordering wave vector
  remains
  commensurate, $Q = Q_0 = \pm 4\pi/3$,
for all $\Delta > 0$, and the
 evolution from quantum-preferred
V state to classically-preferred cone state proceeds differently,
 via two first-order phase transitions  (see Fig. \ref{fig:xxz}).
The V state with $\theta = \ell\pi/3$ survives up to
some critical $\Delta_{c1}
\sim 1/S$, where another commensurate
co-planar order develops, for which $\theta = (2 \ell + 1)\pi/6$. The corresponding spin pattern
  resembles Greek letter $\Psi$ and
  we
  label this state a $\Psi$ phase.
The $\Psi$ phase survives up to  $\Delta_{c2}
 \geq \Delta_{c1}$,
  beyond which the spin configuration turns into
 the commensurate cone state.

 We now discuss the model and the calculations which lead to phase diagrams in
 Figs.~\ref{fig:anis} and ~\ref{fig:xxz}.

{\bf The model.}
 The isotropic Heisenberg antiferromagnet on a triangular lattice is described by  the Hamiltonian
\be
{\cal H}_0 = \frac{1}{2} J\sum_{{\bf r}, {\bm \delta}}  {\bf S}_{\bf r} \cdot {\bf S}_{{\bf r}+{\bm \delta}}  - \sum_{\bf r} h S^z_{\bf r},
\label{eq:ideal}
\ee
where ${\bm \delta}$ are nearest-neighbor vectors of the triangular lattice. The two perturbations we consider are
\begin{eqnarray}
\label{eq:anis}
\delta {\cal H}_{\rm anis} &=& (J' - J) \sum_{\bf r} {\bf S}_{\bf r} \cdot ({\bf S}_{{\bf r} + {\bm \delta}_1} + {\bf S}_{{\bf r} + {\bm \delta}_3}),\\
\delta {\cal H}_{\rm xxz} &=& \frac{1}{2}(J_z - J)\sum_{{\bf r}, \pm {\bm \delta}_{1,2,3}}  S^z_{\bf r} S^z_{{\bf r}+{\bm \delta}} .
\label{eq:xxz}
\end{eqnarray}
where $\langle {\bf r}, {\bf r} + {\bm \delta}_{1,3}\rangle$ are diagonal bonds.

We consider a quasi-classical limit $S \gg1$, when quantum fluctuations are small in $1/S$ and quantum and classical tendencies
compete at small anisotropy $\delta J/J \sim 1/\sqrt{S}$ and/or $\Delta/J \sim 1/S$.
In this limit, the calculations in the vicinity of the saturation field can be done
 using a well-established dilute Bose gas expansion and are controlled by simultaneous smallness of $1/S$ and of
 $(h_{\rm sat} - h)/h_{\rm sat}$~\cite{nikuni,ueda,chen,kamiya}.
 We argue that our results are applicable for all values of $S$, down to $S=1/2$, because (i) quantum selection of the V state
   holds even for $S=1/2$ \cite{chen}, and (ii) numerical analysis of $S=1/2$ systems~\cite{chen,yamamoto}
identified the same phases near saturation field as found here.

We set quantization axis along the field direction and express spin operators ${\bf S}_{\bf r}$
 in terms of Holstein-Primakoff bosons $a, a^+$ as $S^{-}_{\bf r}  =  [2S - a^+_{\bf r} a_{\bf r}]^{1/2} a_{\bf r}^+, ~S^z_{\bf r} = S - a^+_{\bf r} a_{\bf r}$.
 Substituting
  this transformation into ${\cal H}_{\rm anis/xxz}$ and expanding the square root one obtains
the spin-wave Hamiltonian ${\cal H} = {\cal E}_{\rm cl} + \sum_{j=2}^\infty {\cal H}^{(j)}$,
where ${\cal E}_{\rm cl}$ stands for the classical
ground state energy, and
${\cal H}^{(j)}$ are of $j$-th order in operators $a, a^+$.
For our purposes, terms up to $j=6$ have to be retained in the expansion (see the Supplement \cite{suppl}
for technical details).
The
 quadratic part of the spin-wave Hamiltonian reads
\begin{equation}
{\cal H}^{(2)} =  \sum_{\bf k}
 (\omega_{\bf k} - \mu) a_{\bf k}^+ a_{\bf k}
\label{eq:harmonic}
\end{equation}
where $\omega_k = S (J_{\bf k} - J_{{\bf Q}})$ is the spin-wave dispersion, measured relative to its minimum at the saturation field $h_{\rm sat}$, and
$\mu = (h_{\rm sat} - h)/h_{\rm sat}$
 plays the role of chemical potential. For $J-J'$ model,
 $J_{\bf k} = \sum_{\pm \delta_j} J_{\delta_j} (e^{i {\bf k} \cdot \delta_j} - 1)$, where $J_{\delta_{1,3}} = J'$ and $J_{\delta_2} = J$.
Here ${\bf Q} = {\bf Q}_{\rm i} = (Q_{\rm i},0)$ with $Q_{\rm i} = 2 \cos^{-1}(-J'/2J)$.
For XXZ model, $J_{\bf k} = \sum_{\pm \delta_j} (J e^{i {\bf k} \cdot \delta_j} - J_z)$ and ${\bf Q} = {\bf Q}_0 = (4\pi/3,0)$.
In both cases, lowering of a magnetic field below $h_{\rm sat}$  makes $(\omega_{\bf k} - \mu)$ negative at $ {\bf k} \approx \pm {\bf Q}$,
 where ${\bf Q}$ is either ${\bf Q}_{\rm i}$ or ${\bf Q}_0$,
 and drives the Bose-Einstein condensation (BEC) of magnons.
To account for BEC, we introduce
two condensates, $\langle a_{\bf Q}\rangle = \sqrt{N} \psi_1$ and $\langle a_{-\bf Q}\rangle = \sqrt{N} \psi_2$,
 where $\psi_{1,2}$ are complex order parameters.
In real space,
\begin{equation}
\langle a_{\bf r}\rangle = \frac{1}{\sqrt{N}}\sum_{\bf k} e^{i {\bf k} \cdot {\bf r}} \langle a_{\pm {\bf k}}\rangle =
\psi_1 e^{i {\bf Q} \cdot {\bf r}} + \psi_2 e^{- i {\bf Q} \cdot {\bf r}} .
\label{eq:cond}
\end{equation}
The ground state energy, per site, of the {\em uniform} condensed ground state
 is expanded in powers of $\psi_{1,2}$ as
\begin{eqnarray}
&&E_0/N = -\mu (|\psi_1|^2 + |\psi_2|^2) + \frac{1}{2} \Gamma_1(|\psi_1|^4 + |\psi_2|^4) \nonumber\\
&&+\Gamma_2 |\psi_1|^2 |\psi_2|^2 + \Gamma_3 ((\bar{\psi}_1 \psi_2)^3 + {\text{h.c.}}) ...
\label{eq:energy}
\end{eqnarray}
where
$\bar{\psi}_j$ denotes complex conjugated of $\psi_j$, dots stand for higher order terms, and we omitted a constant term.
We verified~\cite{suppl} that higher orders in $\psi_j$
do not
 modify
 our analysis.

Whether the state at $\mu = 0+$ is co-planar or chiral
is decided by the sign of
$\Gamma_1 - \Gamma_2$ \cite{nikuni}. For $\Gamma_1 < \Gamma_2$,
it is energetically favorable to break $Z_2$ symmetry between condensates and choose
$\psi_1  \neq 0, \psi_2  =0$ or vice versa.
Parameterizing the condensate as
$\psi_1 = \sqrt{\rho} e^{i \varphi}$,
where
 $\rho = \mu/\Gamma_1$,
and using Eq.\eqref{eq:cond}, we
 obtain the cone configuration,  Eq.\eqref{eq:umbrella}.  The order parameter manifold of this state is $O(2) \times Z_2$, where $O(2)$ is associated with the phase $\varphi$.

When $\Gamma_1 > \Gamma_2$, it is energetically favorable to preserve $Z_2$ symmetry and  develop both condensates with
 equal magnitude $\rho = \mu/(\Gamma_1+\Gamma_2)$, i.e., set
$\psi_1   = \sqrt{\rho} e^{i \theta_1},
 \psi_2   = \sqrt{\rho} e^{i \theta_2}$.
This corresponds to co-planar  state  with the common phase
$\varphi = (\theta_1 + \theta_2)/2$ and the relative phase $\theta = (\theta_1 - \theta_2)/2$.
The order parameter in this state
  is given by Eq. \eqref{eq:V} with
 ${\bf Q}$ equal to either ${\bf Q}_{\rm i}$ ($J-J'$ model)  or ${\bf Q}_0$ (XXZ model).
 For ${\bf Q} = {\bf Q}_{\rm i}$,
  the state
is incommensurate co-planar
 configuration in Fig. \ref{fig:anis}.
  The order  parameter manifold of this state is
$O(2) \times O(2)$, where one $O(2)$ is associated with $\varphi$ and the other with $\theta$.
For  ${\bf Q} = {\bf Q}_0$,  the co-planar order is commensurate.
 In this case, the symmetry
 is further reduced by $\Gamma_3$ term, which is allowed because
 $e^{i 3 {\bf Q}_0 \cdot {\bf r}} = 1$ for all sites ${\bf r}$ of the lattice.
 This term locks
the relative phase of the condensates $\theta$ to three values,
reducing the broken symmetry
 to $O(2) \times Z_3$.
For $\Gamma_3 < 0$,
$\theta = \pi \ell/3$, where $\ell = 0,1,2$.
For $\Gamma_3 > 0$,
$\theta = (2\ell + 1) \pi/6$.
These are $V$ and
 $\Psi$ states in Figs. \ref{fig:anis} and \ref{fig:xxz}.

Accidental degeneracy of the isotropic  model \eqref{eq:ideal} in the classical limit shows up via $\Gamma_1^{(0)} = \Gamma_2^{(0)} = 9 J$ and
$\Gamma_3^{(0)}  = 0$, where the superscript `0' indicates that these expressions are of zeroth order in $1/S$.
We now analyze the situation in the presence of anisotropy and quantum fluctuations.  We
first consider
$J-J'$ model with $J \neq J'$, and then XXZ model
 with $J_z \neq J$.

{\bf Phases of the $J-J'$ model.}~~~~
We computed $\Gamma^{(0)}_{1,2}$ for classical spins, but in the presence of the the spatial anisotropy
  and found that it
 tilts the balance in favor of the cone phase:
$\Delta \Gamma^{(0)} = \Gamma_2^{(0)} - \Gamma_1^{(0)}  = J (1 - J'/J)^2 (2 + J'/J)^2 >0$.
Quantum $1/S$ corrections, on the other hand, favor the co-planar state: $\Delta \Gamma^{(1)} <0$. We obtained~\cite{suppl}
 \begin{eqnarray}
\Delta \Gamma^{(1)} &=& \frac{1}{16 S} \sum_{{\bf k}\in \text{BZ}} \Big( \frac{(J_0 + 5 J_{\bf k})^2}{J_0 - J_{\bf k}} -
\frac{(J_0 - 4 J_{{\bf Q} + {\bf k}})^2}{J_{{\bf Q} + {\bf k}}  - J_{\bf Q} }\Big) \nonumber\\
&& + \frac{3J}{8S} \approx - \frac{1.6 J}{S}.
 \label{eq:gam12}
 \end{eqnarray}
Combining classical and quantum contributions, we find that
\begin{equation}
\Delta \Gamma = \Delta \Gamma^{(0)} + \Delta \Gamma^{(1)} = \frac{9(\delta J)^2}{J} - \frac{1.6 J}{S}
\label{eq:d-gamma}
\end{equation}
where, we remind, $\delta J \equiv J - J'$.
We see that  $\Delta \Gamma <0$ for $\delta J < \delta J_{c} = 0.42J/\sqrt{S}$, and $\Delta \Gamma >0$ for larger
$\delta J$.
 The condition $\Delta \Gamma =0$ selects the point
 $B$ in Fig. \ref{fig:anis}~\cite{footn}.

 {\it Split transitions near $\delta J_{\rm c}$.}
At $\mu = 0+$, the transition between incommensurate planar and cone phases is first order with no hysteresis.
 We now analyze how this
 transition occurs at a finite positive $\mu \neq 0$.
 We
  depart from the cone state to the right of point B in Fig.~\ref{fig:anis} and move to smaller $\delta J$.
  Suppose that
  the condensate
   in the cone state has momentum
    $+{\bf Q}_{\rm i}$.
  Then
  Goldstone
   spin-wave
   mode is at   ${\bf k}= {\bf Q}_{\rm i}$,
  while
   excitations near ${\bf k}= - {\bf Q}_{\rm i}$ have a finite gap.
 We computed the  excitation spectrum $\omega^{(1)}_{\bf k}$
  with quantum $1/S$ corrections
 and found~\cite{suppl} that  near ${\bf k} \approx - {\bf Q}_{\rm i}$
 \begin{eqnarray}
 \label{c_1}
 &&\omega^{(1)}_{\bf k} \approx \frac{3J}{4} \left[(k_x + {\bar Q}_{\rm i})^2 + k^2_y + \epsilon_{\rm min}\right],\\
 &&\epsilon_{\rm min} = \frac{12 \mu}{h_{\rm sat} J^2} \left[ (\delta J)^2 - (\delta J_{\rm c})^2
 \left(1 + \frac{\mu}{h_{\rm sat}}\right) \right],
 \end{eqnarray}
where ${\bar Q}_{\rm i} = Q_{\rm i}  + (4\pi/3 -Q_{\rm i})(3\mu/h_{\rm sat}) \approx
Q_{\rm i} +1.45 \mu/(h_{\rm sat}\sqrt{S})$.
 The cone state becomes unstable at $\epsilon_{\rm min} =0$, i.e., at $\delta J_{c3} \approx \delta J_{\rm c}(1 + \mu/(2h_{\rm sat}))$,
  and gives rise to magnon condensation with momentum
$(-{\bar Q}_{\rm i}, 0)$, which is {\it different} from $-{\bf Q}_{\rm i}$.
 The condensation of magnons with  $(-{\bar Q}_{\rm i}, 0)$ then gives rise to a secondary cone order,
with momentum not related by symmetry to that of the primary cone order.
 The resulting spin configuration
is a double cone
  with $O(2) \times O(2) \times Z_2$ order parameter manifold.
The
 primary
  condensate sets the transverse component of $\langle {\bf S}^{\perp}_{\bf r}\rangle = \langle S^x_{\bf r} + i S^y_{\bf r}\rangle$ to be
 $\exp[ i {\bf Q}_{\rm i} \cdot {\bf r} + i \theta_1]$
 and  the second condensate
 adds
 $\exp[ - i {\bf {\bar Q}_{\rm i}} \cdot {\bf r} + i \theta_2]$.

 At smaller $\delta J \leq \delta J_{c3}$
 the position of the minimum in $\omega^{(1)}_k$ in (\ref{c_1}) evolves and drifts towards $-{\bf Q}_{\rm i}$.
 Once it reaches $-{\bf Q}_{\rm i}$,
  at $\delta J = \delta J_{c2}$,
 the two cone configurations
 interfere constructively and
give rise to an incommensurate co-planar state.
Critical $\delta J_{c2}$
can be estimated by requiring that $\omega^{(1)}_k =0$ at ${\bf k} = -{\bf Q}_{\rm i}$.
This yields $\delta J_{c2} = \delta J_{c3} (1 - O(\mu/h_{\rm sat})) < \delta J_{c3}$.
We see therefore that the transformation from a cone to an incommensurate co-planar state at at a finite $\mu$ (i.e, at $h \leq h_{\rm sat}$)
occurs via two transitions
at $\delta J_{c2}$ and $\delta J_{c3}$ and involves an intermediate double cone phase (Fig. \ref{fig:anis}).

{\it Instability of the V phase.} We now return to Eq. (\ref{eq:energy}) and consider the transition between the $V$ phase
 and the incommensurate co-planar phase. At $\mu = 0+$, this transition holds at infinitesimally small $\delta J$
  (point A in Fig. \ref{fig:anis}).
 We  show that at a finite $\mu$, the $V$ phase survives up to a
 finite
 $\delta J_{c1} \sim (J/\sqrt{S})  (\mu/h_{\rm sat})$. The argument is
that in the $V$ phase
 ${\bf Q} = {\bf Q}_0$
 is  commensurate
  and $\Gamma_3$ term in Eq. (\ref{eq:energy}) is allowed.
   We recall that at $\delta J=0$ and for classical spins $\Gamma_3=0$.
We
 computed  the classical contribution to $\Gamma_3$ at  $\delta J >0$ and
 the contribution due to quantum fluctuations at $\delta J =0$.  We
 found~\cite{suppl} that  the classical  contribution vanishes, but the quantum contribution is finite to order $1/S^2$  and
 makes $\Gamma_3$ negative:
\begin{eqnarray}
&&
\Gamma_3  = \frac{3}{32 S^2} \sum_{{\bf k}\in \text{BZ}} \Big(\frac{(5 J_{\bf k} + J_0)(5 J_{{\bf Q} + {\bf k}} + J_0) J_{{\bf Q} - {\bf k}}}
{(J_0 - J_{\bf k}) (J_0 - J_{{\bf Q} + {\bf k}})} - \nonumber\\
&& - \frac{(5 J_{\bf k} + J_0) (J_{\bf k} + J_0)}{2 (J_0 - J_{\bf k})} \Big) +  \frac{3J_0}{64 S^2} \approx
 -\frac{0.69 J}{S^2}
\label{eq:gam3}
\end{eqnarray}
Because $\Gamma_3<0$, the $V$ phase has extra negative energy compared to incommensurate phases, and one needs a finite $\delta J$ to overcome this energy difference.

We now argue that the transition at $\delta J_{c1}$ belongs to the special class of  CIT.  To see this, we
  allow for spatially non-uniform configurations of the condensate
$\psi_{1,2}({\bf r})$. This
adds spatial gradient terms to \eqref{eq:anis}:
 the  isotropic term
 ${\cal H}_0$ produces
 conventional quadratic in gradient contribution  $\propto \rho (\partial_x \theta)^2$, while
 $\delta {\cal H}_{\rm anis}$ adds
a {\em linear} gradient term $\propto \rho S \delta J \partial_x (\theta_1 - \theta_2)$.
Combining these two classical contributions with the quantum $\Gamma_3$ term in \eqref{eq:energy},
we obtain
the energy density for the relative phase $\theta = (\theta_1 - \theta_2)/2$:
\begin{equation}
{\cal E}_{\rm \theta} = \frac{3 J S^2 \mu}{4 h_{\rm sat}} (\partial_x \theta)^2 + \frac{\sqrt{3} \delta J S^2 \mu}{h_{\rm sat}} \partial_x \theta +
S \frac{(\Gamma_3 S^2)}{4} \frac{\mu^3}{h_{\rm sat}^3} \cos[6 \theta]
\label{eq:SG}
\end{equation}
Eq.~\eqref{eq:SG} is of standard sine-Gordon form, which allows us to borrow the results from \cite{chen}: the equilibrium value of $\theta$
shifts from the commensurate $\theta = \pi \ell/3$ in the V phase to an
incommensurate value when the coefficient of the linear gradient
term in \eqref{eq:SG}
exceeds the
 geometric
mean of the coefficients of two other terms in \eqref{eq:SG}.
Using Eq. (\ref{eq:SG}) we find that  CIT  occurs at
 $\delta J_{c1} = 1.17 (J/\sqrt{S}) (\mu/h_{\rm sat}) = 0.13 \mu/S^{3/2}$
  (line AC in Fig. \ref{fig:anis}).
 At $\delta J > \delta J_{c1}$, $\theta$ acquires linear dependence on $x$: $\theta = {\tilde Q} x + {\tilde{\theta}}$.
 In this situation,
 the spin configuration becomes
 incommensurate but remains co-planar (Fig. \ref{fig:anis}).

The critical $\delta J_{c1}$ for the CIT has to be compared with $\delta J_{\rm sw}$ at which spin-wave excitations
 in the  V phase soften. We computed spin-wave velocity with quantum $1/S$ corrections and found that it does go down with increasing
 $\delta J$  but vanishes only
  at $\delta J_{\rm sw} \sim (J/\sqrt{S}) (\mu/h_{\rm sat})^{1/2} \gg \delta J_{c1}$.
 This implies that the spin-wave velocity remains finite across the  CIT.

{\bf Phases of ${\cal H}_{\rm xxz}$.}
 For the XXZ model with exchange anisotropy,
  $J$ and $J'$ remain equal,  but $J_z < J_\perp = J$
  on all bonds.
  We verified~\cite{suppl} that ${\bf Q}$
 remains  commensurate  for all $J_z/J \leq 1$, i.e., ${\bf Q} = {\bf Q}_0 = (4\pi/3, 0)$.
In this situation, we found
  $\Gamma_2^{(0)} - \Gamma_1^{(0)}  = - J_{\bf Q} (1 - J_z/J) = 3 J\Delta$.
Quantum corrections to $\Gamma_1$ and $\Gamma_2$ are determined within
 the same isotropic model \eqref{eq:ideal} and are given by
\eqref{eq:d-gamma}. Using this,  we immediately find that the ground state of the quantum
XXZ model is coplanar
for $\Delta \leq \Delta_{c2} =  0.53 /S$ and is a cone for $\Delta > \Delta_{c2}$.
The transition between co-planar and cone states near $\Delta_{c2}$ remains first-order for a finite $\mu > 0$,
 i.e., no intermediate double spiral state appears.
This is the consequence  of the fact that ${\bf Q} = {\bf Q}_0$ remains commensurate.
Still, the transformation from the V phase to the cone phase does involve a new intermediate state, which comes about due to the
change of sign of $\Gamma_3$. Exchange anisotropy $\Delta$
gives rise to a positive
$\Gamma_3$ to order $1/S$: $\Gamma^{(1)}_3 = J (1 + 2 J_z/J) (1 - J_z/J)/(2 S) \approx 3 J \Delta/(2 S)$
 (see \cite{suppl} for details).
At the same time the quantum corrections give rise to {\em negative}
 $\Gamma_3$ to order $1/S^2$ already at $\Delta=0$,  see \eqref{eq:gam3}.
 Combining the two, we
  find
  that
\begin{equation}
\Gamma_3 = \Gamma_3^{(1)} + \Gamma_{3}^{(2)} =  \frac{3 J\Delta}{2 S} - \frac{0.69 J}{S^2} .
\label{eq:A}
\end{equation}
changes sign at $\Delta_{c1} = 0.45 /S < \Delta_{c2} = 0.53/S$.
At smaller $\Delta < \Delta_{c1}$, $\Gamma_3 <0$, and the spin configuration is the V state
(the energy is minimized by setting $\cos 6\theta =1$, see \eqref{eq:energy}).
However, in the interval $\Delta_{c1} < \Delta < \Delta_{c2}$, $\Gamma_3 >0$ becomes  positive.
The energy is now minimized by
$\cos 6 \theta =-1$, which corresponds to
 the $\Psi$ state in Fig. \ref{fig:xxz}.
 The transition is highly unconventional symmetry-wise because the order parameter manifold is  $O(2) \times Z_3$ in both phases,
 but extends to a larger $O(2)\times O(2)$ symmetry at the transition point.

 We present the  phase diagram of XXZ model in Fig. \ref{fig:xxz}.
 A very similar phase diagram has been recently obtained  in the numerical cluster mean-field analysis of the $S=1/2$  XXZ
  model~\cite{yamamoto}.

 To summarize, in this paper we considered anisotropic
 2D Heisenberg antiferromagnets on a triangular lattice  in a high magnetic field close to the saturation. We analyzed the cases of
  spatially anisotropic  interactions, like in Cs$_2$CuCl$_4$ and Cs$_2$CuBr$_4$ and of exchange anisotropy, as in Ba$_3$CoSb$_2$O$_9$.
  We showed that the phase diagram in field/anisotropy plane
  is quite rich due to competition between classical and quantum orders, which favor non-coplanar and co-planar states, respectively.
    This competition leads to multiple transitions and highly non-trivial intermediate phases, including a novel double cone state.
    We demonstrated that one of the transition in each of the two cases studied is of CIT
    type and is not accompanied by softening of spin-wave excitations.

The analysis of this paper can be easily extended to quasi-2D layered systems, with inter-layer antiferromagnetic interaction
 $0 < J'' \ll J$. This additional exchange interaction leads
 to the staggering of coplanar spin configurations, of either V or $\Psi$ kind,
 between the adjacent layers, as can easily be seen by treating $\varphi\to \varphi_z$ in
 Eq.\eqref{eq:V} as layer-dependent variable with discrete index $z$.
 One then immediately finds that $J'' \sum_{{\bf r}, z} \vec{S}_{{\bf r}, z} \cdot \vec{S}_{{\bf r}, z+1}$ is minimized by
$\varphi_z = \varphi + \pi z$,  in
 agreement
  with earlier spin-wave \cite{gekht} and Monte Carlo \cite{koutroulakis} studies.

 We acknowledge useful conversations with L. Balents and C. Batista.
This work is supported by DOE grant DE-FG02-ER46900 (AC) and NSF grant DMR-12-06774 (OAS and WJ).

\begin{widetext}

\renewcommand{\theequation}{A-\arabic{equation}}
  \setcounter{equation}{0}  
  \renewcommand{\thefigure}{A-\arabic{figure}}
   \setcounter{figure}{0}
\newpage
\begin{center}
\bf{Supplementary Information for
`` Phases of triangular lattice antiferromagnet near saturation'' \\ Oleg A. Starykh, Wen Jin, and Andrey V. Chubukov}
\end{center}

Here we present technical details of calculations reported in the manuscript. All calculations were carried out in one-sublattice and in three-sublattice basis,
and led to identical results. For definiteness, we present the details of calculations in the one-sublattice basis.

\section{The Hamiltonian and the expansion in bosons}

We consider  Heisenberg Hamiltonian of 2D triangular lattice (Eq. (3) of the main text), and expand
 it to sixth order in Holstein Primakoff bosons around the ferromagnetic state, which holds at $h > h_{sat}$.   We then move to fields below the saturation value by introducing magnon condensates and using the technique of dilute Bose-gas expansion.

   The Hamiltonian in terms of Holstein Primakoff bosons has the form
\begin{eqnarray}
 {\cal H}& = & {\cal H}^{\rm(2)} + {\cal H}^{\rm(4)} + {\cal H}^{\rm(6)},\nonumber\\
\label{eq:H^2}
 {\cal H}^{\rm(2)} &= &\sum_{\bf k} (\omega_{\bf k} - \mu) a_{\bf   k}^\dag a_{\bf k} ,\\
 \label{eq:H^4}
 {\cal H}^{\rm(4)}  &= &\frac{1}{2N}\sum_{\bf k,k',q}V_{\bf q}({\bf k,k'})   a_{\bf k+q}^\dag a_{\bf k'-q}^\dag a_{\bf k'} a_{\bf k} ,\\
 {\cal H}^{\rm(6)} &= &\frac{1}{\rm 16SN^{2}}\sum_{\bf k, k',k'',q,p}U_{\bf q,p}({\bf k,k',k''}) a_{\bf k+q+p}^\dag a_{\bf k'-q}^\dag a_{\bf k''-p}^\dag a_{\bf k''} a_{\bf k'} a_{\bf k} .
 \label{eq:H^6}
\end{eqnarray}
Here, $a,a^\dag $ are  boson operators, $ \omega_k $ is the magnon dispersion, $ \mu =h_{\rm sat} -h $ is the chemical potential, and
$V_{\bf q}({\bf k,k'}), U_{\bf q,p}(\bf k,k',k'') $ are 2- and 3-body interaction potentials which we list below separately for isotropic and anisotropic models.
Both $\omega_{\bf k}$ and $h_{sat}$ are of order $S$, and we consider $\mu$ also of order $S$.

\subsection{Isotropic Heisenberg Model}
\label{sec:A1}

In the isotropic case
 \begin{eqnarray}
\label{eq:omega}
 \omega_{\bf k} & = & S (J_{\bf k} - J_{\bf Q}),\\
\label{eq:V^4}
  V_{\bf q}({\bf k,k'}) &= &\frac{1}{2}[J_{\bf k-k'+q}+J_{\bf q}-\frac{1}{2}(J_{\bf k+q}+J_{\bf k'-q}+J_{\bf k}+J_{\bf k'})],\\
U_{\bf q,p}(\bf k,k',k'') &= &\frac{1}{9}\Big (J_{\bf k+q}+J_{\bf k''+q}+J_{\bf k+k''-k'+q}+J_{\bf k+p}+J_{\bf k'+p}+J_{\bf k+k'-k''+p}\nonumber\\
    && +J_{\bf k'+k''-k-q-p} +J_{\bf k''-q-p}+J_{\bf k''-q-p}\Big)\nonumber\\
   && -\frac{1}{6}\Big(J_{\bf k+q+p}+J_{\bf k'-q}+J_{\bf k''-p}+J_{\bf k}+J_{\bf k'}+J_{\bf k''}\Big).
 \label{eq: U^6}
\end{eqnarray}
where $J_{\bf k} = 2J( \cos[k_x] + 2 \cos[\frac{k_x}{2}] \cos[\frac{\sqrt{3} k_y}{2}]),$ with its minimum $ J_{\bf Q}$ at $ {\bf Q}=(Q_0,0),$ and $Q_0=4\pi/3.$

\subsection{Anisotropic $J$-$J'$ Model}

In this model, $\omega_{\bf k}$,  $V_{\bf q}({\bf k,k'})$, and $U_{\bf q,p}({\bf k,k',k''})$ are all in the same form as $J_{\bf k}$ above, except replacing all $J_{\bf k}$ with
$\tilde J_{\bf k}$, where $\tilde{J}_{\bf k} = 2(J \cos[k_x] + 2 J'\cos[\frac{k_x}{2}] \cos[\frac{\sqrt{3} k_y}{2}])$. $\tilde{J}_{\bf k}$  has minimum $\tilde J_{\bf Q}$  at
$ {\bf Q}=(Q_{\rm i},0)$, and $ Q_{\rm i}=2\cos^{-1}[-J'/2J]$.

\subsection{ XXZ Model}

In this model, $\omega_{\bf k}$ is same as Eq.\eqref{eq:omega}, and  $U_{\bf q,p}(\bf k,k',k'')$ is same as  Eq.\eqref{eq: U^6}.
The difference comes from  $V_{\bf q}(\bf k,k')$, which now contains the exchange anisotropy in the $z$ direction:
\begin{equation}
V_{{\bf q}}({\bf k,k'})=\frac{1}{2}\Big[J^{ z}_{\bf k-k'+q}+J^{z}_{\bf q}-\frac{1}{2}(J_{\bf k+q}+J_{\bf k'-q}+J_{\bf k}+J_{\bf k'})\Big],
\end{equation}
where $J^{z}_{\bf k} = 2J^{z}( \cos[k_x] + 2 \cos[\frac{k_x}{2}] \cos[\frac{\sqrt{3} k_y}{2}])$. The minimum of $J^{z}_{\bf k}$  is at  $ {\bf k}=(Q_0,0).$

\section{ Calculation of  $\Gamma_1,\Gamma_2,\Gamma_3$}

We follow~\cite{griset} and split magnon operators into condensate and non-condensate fractions as
\begin {equation}
a_{\bf k}=\sqrt{N}\psi_{1}\delta_{\bf k,Q}+\sqrt{N}\psi_{2}\delta_{\bf k,-Q}+\tilde a_{\bf k},
\label{eq:gs}
\end {equation}
where $\psi_{1,2}$ describe condensates at momenta ${\bf k} = {\bf Q}$ and ${\bf k} = -{\bf Q}$,
 and $\tilde a_{\bf k}$ describes non-condensate magnons. The ground state energy density reads
\begin{eqnarray}
E_0/N= -\mu (|\psi_1|^2 + |\psi_2|^2) + \frac{1}{2} \Gamma_1(|\psi_1|^4 + |\psi_2|^4)
+\Gamma_2 |\psi_1|^2 |\psi_2|^2 + \Gamma_3 ((\bar{\psi}_1 \psi_2)^3 +\rm\bf h.c.)
\end{eqnarray}
The classical expressions  for  $\Gamma_1$ and $  \Gamma_2$ (the ones at order $1/S^0$) are obtained by
neglecting all non-condensate modes and are shown schematically in
Fig.\ref{fig:A1}.  These contributions are related to potential $V_{\bf q}({\bf k, k'})$ via
\begin{eqnarray}
&& \Gamma_{1}^{(0)}= V_{\bf 0}(\bf Q,Q),\\
&& \Gamma_{2}^{(0)}= V_{\bf 0}(\bf Q,-Q)+V_{\rm 2\bf Q}(-Q,Q).
\end{eqnarray}

The classical expression  for  $\Gamma_3$ (at order $1/S$) is shown schematically in Fig.\ref{fig:A1}
and it is related to potential $V_{\bf q}({\bf k, k'})$ and $U_{\bf q,p}({\bf k, k', k''})$ via
\begin{equation}
\Gamma_{3}^{(1)}= \frac{U_{2\bf Q,\rm 2\bf Q}(\bf Q,Q,Q)}{16S}-\frac {[\rm V_{2\bf Q}(\bf Q,Q)]^2}{\omega_{\rm 3\bf Q}}.
\label{eq:gamma3}
\end{equation}
Here the first term comes directly from the Hamiltonian \eqref{eq:H^6}, and the second one originates from the condensate $\psi_0 \equiv \langle \tilde a_{0}\rangle \ne 0$,
which is induced at the momentum ${\bf k} = 3 {\bf Q} = 0$ in the case of {\em commensurate} ordering at wave vector $ {\bf Q}=(4\pi/3_0,0)$.
This novel condensate adds the term $|\psi_{0}|^2 \omega_{\bf 0}+V_{2\bf Q}(\bf Q,Q)[\psi_{\bf 0}(\bar \psi_{\rm1} \rm \psi_{2}^ {2}+\psi_{1} ^{2}\bar \psi_{2})+\bf h.c]$ to the ground state energy. Minimizing this extra energy contribution, we find the expression for $\psi_{0}$
\begin{equation}
\psi_{0}=-\frac{V_{2\bf Q}(\bf Q,Q)}{ \omega_{0}}(\bar \psi_{\rm1} \rm \psi_{2}^ {2}+\psi_{1} ^{2}\bar \psi_{2})=\rm \frac{1}{\rm 4S}
(\bar \psi_{\rm1} \rm \psi_{2}^ {2}+\psi_{1} ^{2}\bar \psi_{2}).
\label{eq:psi 00}
\end{equation}
It is important to keep in mind that this result is derived for $ {\bf Q}=(4\pi/3_0,0)$, when $e^{i 3 {\bf Q} \cdot {\bf r}} = 1$ for all sites of the
triangular lattice ${\bf r}$.

\begin{figure}[htp]
\includegraphics[scale=0.25]{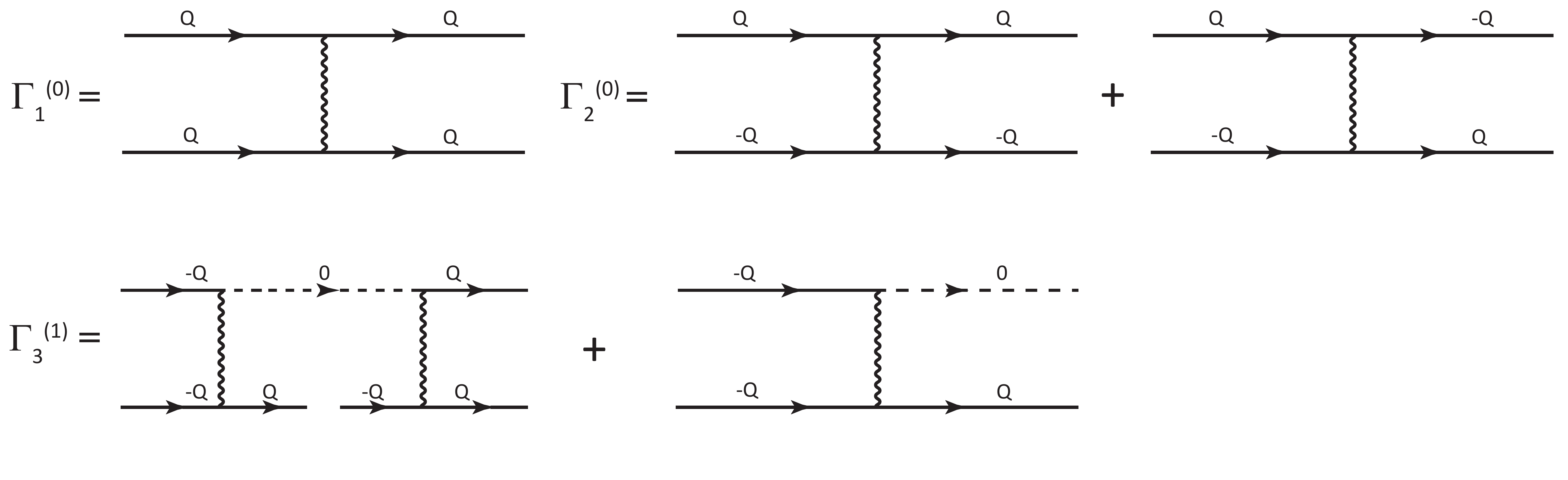}
\caption{Diagrams for $\Gamma_1$, $\Gamma_2 $ and $\Gamma_3$ in the classical limit. }
\label{fig:A1}
\end{figure}

The expressions for $\Gamma_{1}^{(0)}, \Gamma_{2}^{(0)}$, and $\Gamma_{3}^{(1)}$
  are different in the isotropic case and in the two anisotropic cases.\\

For the isotropic model,
\begin{eqnarray}
&& \Gamma_{1}^{(0)}= J_{\bf 0}-J_{\bf Q}, \Gamma_{2}^{(0)}= J_{\bf 0}+J_{2\bf Q}-J_{\bf Q},\nonumber\\
&& \Gamma_{3}^{(1)}=0.
\end{eqnarray}

For $J-J'$ model,
\begin{eqnarray}
&&\Gamma_{2}^{(0)}-\Gamma_{1}^{(0)}= \tilde J_{2\bf Q}-\tilde J_{\bf Q}=J(2+ \frac{J'}{J})^2 (1- \frac{J'}{J})^2 \approx \frac{9(\delta J)^2}{J} ,\nonumber\\
&& \Gamma_{3}^{(1)}= 0.
\end{eqnarray}

For XXZ model,
\begin{eqnarray}
&&\Gamma_{2}^{(0)}-\Gamma_{1}^{(0)}= J^{z}_{2 \bf Q}-J_{\bf Q}=3J\Delta,\nonumber\\
&&\Gamma_{3}^{(1)}=\frac{J_{\bf 0}-J_{\bf Q}}{16S}-\frac {(4 J^z_{\bf Q}-3J_{\bf Q}-J_{\bf 0})^2}{16S(J_{\bf 0} -J_{\bf Q})}
=\frac{J}{2S}(1+\frac{J_z}{J})(1-\frac{J_z}{J})\approx \frac{3J\Delta}{2S}.
\end{eqnarray}

\section {Quantum corrections to $\Gamma_1,\Gamma_2,\Gamma_3$}

In this section, we compute quantum corrections to $\Gamma_1,\Gamma_2,\Gamma_3$. Because these corrections already contain extra factor of $1/S$, they can be
calculated by neglecting anisotropy. Quantum corrections to $\Gamma_{1,2}$ are of order $1/S$, and quantum corrections to $\Gamma_3$ are of order $(1/S)^2$.
In both cases, quantum term has extra factor $1/S$ compared to classical results.
Each quantum correction is a sum of the two terms -- one comes from normal ordering of Holstein-Primakoff bosons, and the other from second and third-order terms in the perturbation expansion in $1/S$.

\subsection{Corrections from normal ordering}

The Holstein-Primakoff transformation
\begin{equation}
S^z (r) = S - a^+_{\bf r} a_{\bf r},  S^{+} = \sqrt{2S - a^+_{\bf r} a_{\bf r}} a_{\bf r},  S^{-} = \sqrt{2S} a^+_{\bf r} \sqrt{2S - a^+_{\bf r} a_{\bf r}}
\end{equation}
contains the square-root $\sqrt{2S - a^+_{\bf r} a_{\bf r}}$, which needs to be expanded in the {\em normal-ordered} form to perform dilute gas analysis (all $a^+_{\bf r}$ have to stand to the left of $a_{\bf r}$).  Because $a^+_{\bf r} a_{\bf r} = a_{\bf r} a^+_{\bf r}-1$, i.e., $(a^+_{\bf r} a_{\bf r})^2 = a^+_{\bf r} a^+_{\bf r} a_{\bf r} a_{\bf r} + a^+_{\bf r} a_{\bf r}$, etc, the prefactors in this {\em normal-ordering} are not simply powers of $1/S$ but rather  contain series of $1/S$ terms. To order $1/S^3$ we have
\begin{eqnarray}
\label{eq:HP}
S^{-}_{\bf r}  =  \sqrt{2S} a_{\bf r}^+ \Big\{1 - \frac{1}{4S}(1 + \frac{1}{8S} + \frac{1}{32S^2}) a^+_{\bf r} a_{\bf r} -\frac{1}{32 S^2} (1 + \frac{3}{4S}) a^+_{\bf r} a^+_{\bf r} a_{\bf r} a_{\bf r} -
\frac{a^+_{\bf r} a^+_{\bf r} a^+_{\bf r} a_{\bf r} a_{\bf r} a_{\bf r}}{128 S^3} + O(1/S^{4})\Big\}\nonumber
\end{eqnarray}
The $1/S$ corrections to the prefactors modify   Eqs.\eqref{eq:H^4} and \eqref{eq:H^6} to
\begin{eqnarray}
\label{eq:delta H4}
&&\delta{\cal H}^{\rm(4)}=-\frac{J}{32S}\sum_{\bf r,\bf \delta}( a_{\bf r}^\dag a_{\bf r}^\dag a_{\bf r} a_{\bf r+\bf \delta}+\bf h.c),\\
&&\delta{\cal H}^{\rm(6)}=\frac{J}{128S^2}\sum_{\bf r,\bf \delta}( a_{\bf r}^\dag a_{\bf{ r+\bf \delta}}^\dag  a_{\bf r+\delta}^\dag a_{\bf r} a_{\bf r} a_{\bf r+\delta}+{\bf h.c})-\frac{3J}{128S^2}\sum_{\bf r,\delta}( a_{\bf r}^\dag a_{\bf r}^\dag  a_{\bf r+\delta}^\dag a_{\bf r} a_{\bf r} a_{\bf r}+\bf h.c).
\label{eq:delta H6}
\end{eqnarray}
Substituting the form of the condensate in real space
 \begin{equation}
\langle a_{\bf r}\rangle = \frac{1}{\sqrt{N}}\sum_{\bf k} e^{i {\bf k} \cdot {\bf r}} \langle a_{\pm {\bf Q}}\rangle =
\psi_1 e^{i {\bf Q} \cdot {\bf r}} + \psi_2 e^{- i {\bf Q} \cdot {\bf r}} .
\label{eq:cond}
\end{equation}
we obtain $1/S$ corrections to classical expressions for  $\Gamma_{1,2,3}$:
\begin{eqnarray}
&&\Delta \Gamma_{a}^{(1)}=\Gamma_{2a}^{(1)}-\Gamma_{1a}^{(1)}=(-\frac{J_{\bf Q}}{4S})-(-\frac{J_{\bf Q}}{8S})=\frac{3J}{8S},\nonumber\\
&&\Gamma_{3a}^{(2)}=\frac{5J_{\bf 0}}{128S^2}+\frac{J_{\bf 0}}{128S^2}=\frac{9J}{32S^2}.
\label{eq:delta gam3}
\end{eqnarray}

\subsection{Corrections from quantum fluctuations}

To find quantum corrections to parameters $\Gamma_{1,2,3}$, we evaluate corrections to the ground state energy density
$\delta E$  from non condensed
modes $\tilde{a}_{\bf k}$ in \eqref{eq:gs} in perturbation theory up to third order and obtain the correction to the ground state energy density $\Delta E$ to sixth order in the condensates $\psi_1$ and $\psi_2$. The prefactors for the $\psi^4$ and $\psi^6$ term in $\Delta E$ yield quantum corrections to interaction parameters $\Gamma_{1,2,3}$ .

Quite generally, under perturbation $H_i$, the partition function is
\begin{eqnarray}
Z= \int \prod_{\bf k} da_{\bf k}^{\dag} da_{\bf k} e^{\int_0^\beta d\tau (L_0 - H_i)}= Z_0 \frac{ \int \prod_{\bf k} da_{\bf k}^{\dag} da_{\bf k}
e^{\int_0^\beta d\tau (L_0 - H_i)}}{ \int \prod_{\bf k} da_{\bf k}^{\dag} da_{\bf k}e^{\int_0^\beta L_0}} \equiv Z_0\langle e^{-\int_0^\beta H_i}\rangle_0.
\end{eqnarray}
Here $L_0 = \sum_{k} (a_k^\dag \frac{\partial }{\partial \tau}a_k) - {\cal H}^{(2)}$ represents Lagrangian of non-interacting magnons described
by the quadratic Hamiltonian \eqref{eq:harmonic}, and $\beta = 1/T$.
The internal energy density is
\begin{eqnarray}
&&E=-\frac{\partial \ln Z}{\partial \beta}\approx-\frac{\partial \ln Z_0}{\partial \beta}-\frac{\partial (\beta \ln \langle e^{-H_i}\rangle)}{\partial \beta} =E_0+ \Delta E
\end{eqnarray}
The correction term $\Delta E$  is represented
by the standard cumulant expansion, which involves only connected averages of the perturbation $H_i$
\begin{equation}
\Delta E= \langle H_{i}\rangle_{0}-\frac{1}{2!}\langle \int_\tau H_{i}^{2}\rangle_{0}+\frac{1}{3!}\langle \int_\tau \int_{\tau'} H_{i}^{3}\rangle_{0}+\dots.
\end{equation}
 In the the zero-temperature limit, in which all our calculations are done, $E = E_0 + \Delta E$  determines the ground state energy.
Integration over relative times $\tau, \tau' \dots$ ensures conservation of frequencies in the internal vertices of the diagrams. The role of the
perturbation $H_i$ is played by interacting Hamiltonians \eqref{eq:H^4}, \eqref{eq:H^6} expressed in terms of condensates $\psi_{1,2}$ and
non-condensed magnons $\tilde{a}_{\bf k}$ after the substitution \eqref{eq:gs}. We remind that the averaging is
over the free-boson Hamiltonian for isotropic system at $h = h_{sat}$.

\subsubsection{Quantum corrections to $\Gamma_{1,2}$}

Quantum corrections to  $\Gamma_{1,2}$ al of order $1/S$, and to get them  we only need  the fourth-order term in bosons \eqref{eq:H^4}:
\begin{equation}
{\cal H_ {\rm i,\bf k}}=\sum_{\bf k}\Big[\Big(  \frac{1}{2}V_{\bf {k}}({\bf Q,Q})\psi_{1}^{2}a_{\bf Q+k}^{\dag} a_{\bf Q-k}^{\dag}+V_{\bf k}({\bf Q,-Q})\psi_{1}\psi_{2}a_{\bf Q+k}^{\dag} a_{\bf -Q-k}^{\dag}+  \frac{1}{2}V_{\bf k}({\bf -Q,-Q})\psi_{2}^{2}a_{\bf -Q+k}^{\dag} a_{\bf -Q-k}^{\dag}\Big)+\bf h.c.\Big],
\label{eq:os2}
\end{equation}
where $V_{\bf q}(k,k')$ is defined in Eq.\eqref{eq:V^4}.
The first-order correction to the energy density obviously vanishes, and the second-order perturbative correction yields
\begin{eqnarray}
\Delta E&=&-\frac{1}{2}\sum_{\bf k,q}\langle{\cal H_ {\rm i,\bf k}}\cdot {\cal H_ {\rm i,\bf q}}\rangle_0\nonumber\\
            &=&-\sum_{\bf k,q}\Big[\frac{1}{4}|\psi_{1}|^{4}V_{\bf k}({\bf Q,Q})V_{\bf q}({\bf Q,Q})\langle a_{\bf Q+k}^{\dag} a_{\bf Q-k}^{\dag}a_{\bf Q+q} a_{\bf Q-q}\rangle_0\nonumber\\
 &\hspace{8mm}& +\frac{1}{4}|\psi_{2}|^{4}V_{\bf k}({\bf -Q,-Q})V_{\bf q}({\bf -Q,-Q})\langle a_{\bf -Q+k}^{\dag} a_{\bf -Q-k}^{\dag}a_{\bf -Q+q}a_{\bf -Q-q}{\rangle_{0}}\nonumber \\
            &\hspace{8mm}& + \frac{1}{2}|\psi_1|^{2}|\psi_2|^{2}V_{\bf k}({\bf Q,-Q})V_{\bf q}({\bf Q,-Q})\langle a_{\bf Q+k}^{\dag} a_{\bf -Q-k}^{\dag}a_{\bf Q+q}a_{\bf -Q-q}\rangle_0\Big].
\end{eqnarray}
By Wick's theorem,
\begin{equation}
\langle a_{k_1}^{\dag}a_{k_2}^{\dag}a_{k_3}a_{k_4}\rangle_0=\langle a_{k_1}^{\dag} a_{k_3}\rangle_0\langle a_{k_2}^{\dag} a_{k_4}\rangle_0+\langle a_{k_1}^{\dag} a_{k_4}\rangle_0\langle a_{k_2}^{\dag} a_{k_3}\rangle_0.
\label{Wick}
\end{equation}
where the pair average is~\cite{popov}
\begin{equation}
\langle a_{k_1}^{\dag} a_{k_2}\rangle_{0} =-\delta_{k_{1},k_{2}}G_{0}(k_1),
\label{pair aver}
\end{equation}
and $G_{0}(k) \equiv G_{0}(k, \epsilon)$ is the free boson Green's function
\begin{equation}
G_{0}(k)=(i\omega - \epsilon_k)^{-1},
\end{equation}
Utilizing the properties of \eqref{Wick} and \eqref{pair aver}, we obtain the terms in the form
\begin{equation}
\sum_{\bf k,q}V_{\bf k}({\bf Q,Q})V_{\bf q}({\bf Q,Q})\langle a_{\bf Q+k}^{\dag} a_{\bf Q-k}^{\dag}a_{\bf Q+q} a_{\bf Q-q}\rangle_0=\sum_{\bf k,\rm \omega}\frac{2V_{\bf k}^{2}({\bf Q,Q})}{(i \omega - \epsilon_{\bf Q+k})(i \omega - \epsilon_{\bf Q-k})}
\end{equation}
Using
\begin{equation}
 ~{\text{with}}~\sum_{\rm \omega} \frac{1}{(i \omega - \epsilon_{1})(i \omega - \epsilon_{2})}=\int \frac{d\omega}{2\pi} \frac{1}{(i \omega - \epsilon_{1})(i \omega - \epsilon_{2})}=\frac{1}{\epsilon_{1}+\epsilon_{2}}
\end{equation}
 and collecting prefactors we obtain the corrections to $ \Gamma_{1,2}$ in the form
\begin{eqnarray}
&& \Gamma_{1b}^{(1)}=-\sum_{\bf k}\frac{V_{\bf k}^{2}({\bf Q,Q})}{\omega_{\bf Q+k}+\omega_{\bf Q-k}}=-\frac{1}{16S}\sum_{\bf k}\frac{(J_{\bf 0}+5J_{\bf k})^2}{J_{\bf 0}-J_{\bf k}},\nonumber\\
&&\Gamma_{2b}^{(1)}=-\sum_{\bf k}\frac{V_{\bf k}^{2}({\bf Q,-Q})}{\omega_{\bf Q+k}}=-\frac{1}{16S}\sum_{\bf k}\frac{(J_{\bf 0}-4J_{\bf Q+k})^2}{J_{\bf Q+k}-J_{\bf k}}.
\label{eq:gamma12-quantum}
\end{eqnarray}
These corrections can be equally obtained diagrammatically, by evaluating second-order corrections to $\phi^4$ vertices, as in Fig. \ref{fig:A2}.
\begin{figure}[b]
\begin{center}
\includegraphics[scale=0.25]{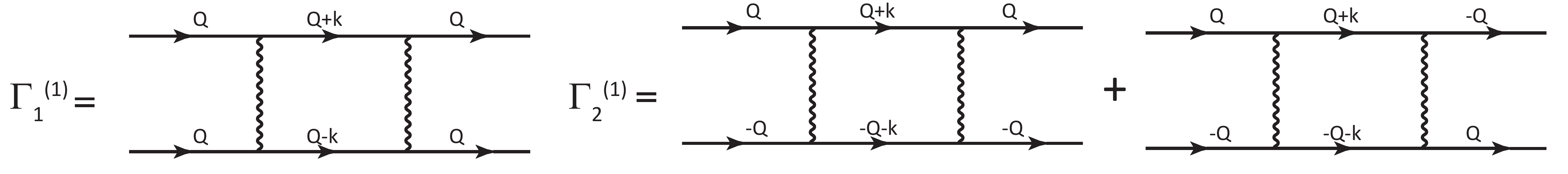}
\caption{Diagrammatic representation of perturbative corrections to $\Gamma_1$ and $\Gamma_2 $.}
\label{fig:A2}
\end{center}
\end{figure}

Each of the two integrals above is logarithmically divergent, but these divergences cancel out in their difference, resulting in a finite result
\begin{equation}
\Delta \Gamma_{b}^{(1)}=\Gamma_{2b}^{(1)}- \Gamma_{1b}^{(1)}=-\frac{1.97J}{S},
\end{equation}
Adding $\Delta \Gamma_{a}^{(1)}$, Eq.\eqref{eq:delta gam3}, to this result we obtain the total quantum correction
$\Delta \Gamma^{(1)} = \Delta \Gamma_{a}^{(1)} + \Delta \Gamma_{b}^{(1)} = -1.595 J/S \approx -1.6 J/S$, as quoted in Eq.\eqref{eq:d-gamma} of the main text.

\subsubsection{Quantum corrections to $\Gamma_{3}$}

Correction to $\Gamma_{3}$ is in order of $(1/S)^2$, and to get such term  in the ground state energy density we need to incude both four-boson and six-boson terms in the Hamiltonian, Eqs. \eqref{eq:H^4} and \eqref{eq:H^6}. We have
\begin{eqnarray}
\cal H_ {\rm i}^{(\rm 4)}&=&\frac{1}{8}\sum_{\bf k}(5J_{\bf k}-2J_{\bf Q})\Big[(\bar{\psi_{1}^{2}}a_{\bf Q+k}a_{\bf Q-k}+\bar{\psi_{2}^{2}}a_{\bf -Q+k} a_{\bf -Q-k})+\bf h.c.\Big]\nonumber\\
                                       &&-\frac{1}{4}\sum_{\bf k}(J_{\bf k}-J_{\bf Q})\Big[(\psi_{0}\psi_{2}a_{\bf Q+k}^{\dag} a_{\bf Q-k}^{\dag}+\psi_{0}\psi_{1}a_{\bf -Q+k}^{\dag} a_{\bf -Q-k}^{\dag})+\bf h.c.\Big],\\
\cal H_ {\rm i}^{(\rm 6)}&=&\frac{1}{16S}\sum_{\bf k}(\frac{5}{2}J_{\bf k}-4J_{\bf Q})\Big[(\bar{\psi_{1}}\psi_{2}^{3}a_{\bf Q+k}^{\dag} a_{\bf Q-k}^{\dag}+\psi_{1}^{3}\bar{\psi_{2}}a_{\bf -Q+k}^{\dag} a_{\bf -Q-k}^{\dag})+\bf h.c.\Big].
\end{eqnarray}
We use the expression of $\psi_0$ in Eq.\eqref{eq:psi 00}, to rewrite $\cal H_ {\rm i}^{(\rm 4)}$ as,
\begin{eqnarray}
\cal H_ {\rm i}^{(\rm 4)}&=& \frac{1}{8}\sum_{\bf k}(5J_{\bf k}-2J_{\bf Q})\Big[(\bar{\psi_{1}^{2}}a_{\bf Q+k}a_{\bf Q-k}+\bar{\psi_{2}^{2}}a_{\bf -Q+k} a_{\bf -Q-k})+\bf h.c.\Big]\nonumber\\
                                       &&-\frac{1}{16S}\sum_{\bf k}(J_{\bf k}-J_{\bf Q})\Big[(\bar{\psi_{1}}\psi_{2}^{3}a_{\bf Q+k}^{\dag} a_{\bf Q-k}^{\dag}+\psi_{1}^{3}\bar{\psi_{2}}a_{\bf -Q+k}^{\dag} a_{\bf -Q-k}^{\dag})+\bf h.c.\Big].
\end{eqnarray}
The total perturbation Hamiltonian is now
\begin{eqnarray}
\cal H_ {\rm i,\bf k}&=&\cal H_ {\rm i}^{(\rm 4)}+\cal H_ {\rm i}^{(\rm 6)}\nonumber\\
&=&\frac{1}{8}\sum_{\bf k}(5J_{\bf k}-2J_{\bf Q})\Big[(\bar{\psi_{1}^{2}}a_{\bf Q+k}a_{\bf Q-k}+\bar{\psi_{2}^{2}}a_{\bf -Q+k} a_{\bf -Q-k})+{\bf h.c.}\Big]\nonumber\\
                                       &&-\frac{3}{32S}\sum_{\bf k}(J_{\bf k}-2J_{\bf Q})\Big[(\bar{\psi_{1}}\psi_{2}^{3}a_{\bf Q+k}^{\dag} a_{\bf Q-k}^{\dag}+\psi_{1}^{3}\bar{\psi_{2}}a_{\bf -Q+k}^{\dag} a_{\bf -Q-k}^{\dag})+\bf h.c.\Big],
                                       \label{ac_1}
\end{eqnarray}
Because of two terms in (\ref{ac_1}), there are two contributions to $\Delta E$ to order $\psi^6/S^2$. One comes from taking the product of $\psi^2$ and $\psi^4$ terms in the second-order perturbation theory.  This yields
\begin{eqnarray}
\Delta E_a&=&-\frac{1}{2}\sum_{\bf k,q}\langle{\cal H_ {\rm i,\bf k}}\cdot {\cal H_ {\rm i,\bf q}}\rangle_0 =
-\frac{3}{128S}\sum_{\bf k,q}(5J_{\bf k}-2J_{\bf Q})(J_{\bf q}-2J_{\bf Q})\times\nonumber\\
             &\times&
             \Big[\psi_{1}^{3}\bar{\psi_{2}^{3}}\langle a_{\bf Q+k}^{\dag} a_{\bf Q-k}^{\dag}a_{\bf Q+q} a_{\bf Q-q}\rangle_{0}+\bar{\psi_{1}^{3}}\psi_{2}^{3}\langle a_{\bf -Q+k}^{\dag} a_{\bf -Q-k}^{\dag}a_{\bf -Q+q} a_{\bf -Q-q}\rangle_{0}\Big]
\label{ac_2}
\end{eqnarray}
and
\begin{equation}
\Delta \Gamma_{3,a}^{(2)}=-\frac{3}{64S^2}\sum_{\bf k}\frac{(5J_{\bf k}-2J_{\bf Q})(J_{\bf k}-2J_{\bf Q})}{J_{\bf 0}-J_{\bf k}}.
\label{eq:gam3.1}
\end{equation}
Diagrammatically, this correction to $\Gamma_{3}$ is given by the first two diagrams in Fig.\ref{fig:A3},

\begin{figure}[b]
\begin{center}
\includegraphics[scale=0.25]{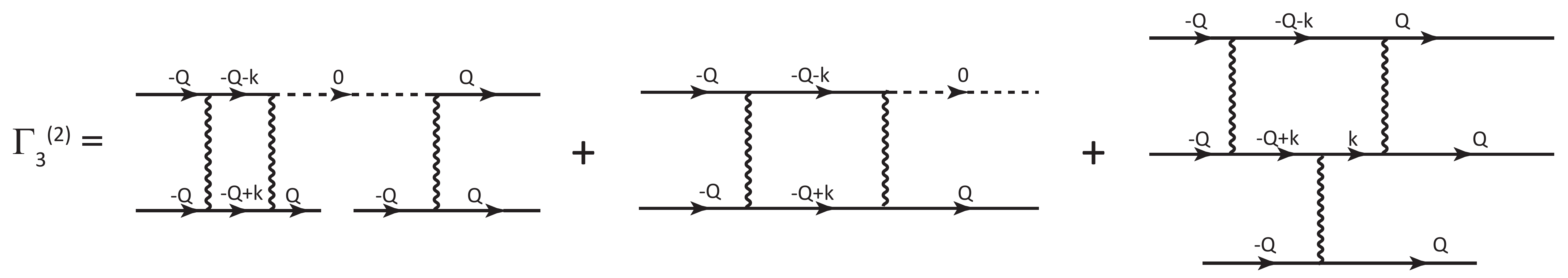}
\caption{Diagrams for $1/S$ corrections to $\Gamma_3 $. The first two diagrams are 2nd order perturbation corrections from the product of $\psi^2$ 
and $\psi^4$ terms in Eq (\ref{ac_1}), the last diagram is 3th order perturbative correction from \eqref{eq:A33}.}
\label{fig:A3}
\end{center}
\end{figure}

Another contribution to $\Delta E$ of order $\psi^6/S^3$ comes  from taking $\psi^2$ term in (\ref{ac_1}) to 3rd order in perturbation theory. The corresponding term in the perturbative Hamiltonian (\ref{ac_1}) comes from fourth-order term in Holstein-Primakoff bosons and we write it separately:
\begin{eqnarray}
\cal H_ {\rm i}^{(\rm 4)}&=&\sum_{\bf k}\Big[\frac{1}{8}(5J_{\bf k}-2J_{\bf Q})(\psi_{1}^{2}a_{\bf Q+k}^{\dag}a_{\bf Q-k}^{\dag}+\bar{\psi_{2}^{2}}a_{\bf -Q+k} a_{\bf -Q-k})+\bf h.c.\Big]\nonumber\\
           &&+\sum_{\bf k}\frac{3}{2}J_{\bf Q-k}(\psi_{1}{\bar{\psi_{2}}}a_{\bf k}^{\dag}a_{\bf Q+k}+\bf h.c.).
           \label{eq:A33}
\end{eqnarray}
The third-order perturbative correction to the ground state density is
\begin{eqnarray}
\Delta E_b&=&\frac{1}{3!}\sum_{\bf k,q,l}\langle{\cal H_ {\rm i,\bf k}}\cdot {\cal H_ {\rm i,\bf q}}\cdot {\cal H_ {\rm i,\bf l}}\rangle_{0}\nonumber\\
  &&= \frac{3}{128}\sum_{\bf k,q,l}\frac{3}{2}J_{\bf Q-k}(5J_{\bf q}-2J_{\bf Q})(5J_{\bf l}-2J_{\bf Q})(\psi_{1}^{3}\bar{\psi_{2}^{3}}+{\bf h.c.})\langle a_{\bf k}^{\dag} a_{\bf Q+q}^{\dag} a_{\bf Q-q}^{\dag}a_{\bf Q+k}a_{\bf -Q+l}a_{\bf -Q-l}\rangle_{0}
\end{eqnarray}
This leads to second $1/S^2$ contribution to  $\Gamma_3$ in the form
\begin{eqnarray}
\Gamma_{3b}^{(2)} &=& \frac{3}{32 S^2} \sum_{{\bf k}} \frac{J_{\bf Q-k}(5 J_{\bf k} + J_0)(5 J_{{\bf Q} + {\bf k}} + J_0) }
{(J_0 - J_{\bf k}) (J_0 - J_{{\bf Q} + {\bf k}})} .
\label{eq:gam3.2}
\end{eqnarray}
In diagrammatic approach, this correction comes from the third diagram in Fig.\ref{fig:A3},

The total $\Gamma_{3}^{(2)}$ is the sum of terms in Eqs.\eqref{eq:gam3.1} and Eq.\eqref{eq:gam3.2}
\begin{eqnarray}
\Gamma_{3}^{(2)} &=& \frac{3}{32 S^2} \sum_{{\bf k}} \Big(\frac{J_{\bf Q - k}(5 J_{\bf k} + J_0)(5 J_{{\bf Q} + {\bf k}} + J_0)}
{(J_0 - J_{\bf k}) (J_0 - J_{{\bf Q} + {\bf k}})} -  \frac{(5 J_{\bf k} + J_0) (J_{\bf k} + J_0)}{2 (J_0 - J_{\bf k})} \Big) = -\frac{0.97J}{ S^2}.
\label{eq:gam3}
\end{eqnarray}
Here again we observe the cancellation of logarithmic singularities, present in the individual integrals.

\section{ Intermediate double cone state for $J-J'$ model}

In this Section, we analyze the phase transition from the cone to the coplanar state, when magnetic field $ h$ is below $h_{\rm sat}$,
i.e., $\mu=h_{\rm sat}-h$ is positive. We remind that at $\mu = 0+$, the cone state is stable at $\delta J = J-J' > \delta J_{c}=0.42J/\sqrt S$.
 Accordingly, we treat $\delta J \approx \delta J_c$ as a small parameter.

Our goal will be to obtain the spin-wave spectrum in the cone state to leading order in $\delta J$ and with quantum corrections.
The magnon modes in the cone state are
\begin{equation}
a_{\bf k}=\sqrt{N}\psi_{1}\delta_{\bf k,Q}+\tilde a_{\bf k}.
\label{eq:cone-a}
\end{equation}
 where, we remind, ${\tilde a}_{\bf k}$ describe non-condensed bosons and $\psi_{1} \propto \sqrt{S}$ describes the condensate fraction.

 We first consider classical spin-wave excitations at the leading order in $1/S$, but a non-zero $\delta J$, and then add quantum $1/S$ corrections to the excitation spectrum.
 As before, the latter already contain $1/S$ and can be computed in the isotropic $\delta J = 0$ limit.

\subsection{Classical spin-wave excitations}

Spatially anisotropic Hamiltonian to second order in ${\tilde a}_{\bf k}$ reads
\begin{eqnarray}
{\cal H}_{\rm anis}&=&H_1+H_2\nonumber\\
\label{H1}
H_1&=&{\cal H}_{\rm anis}^{(2)}=\sum_{\bf k} \Big[S(\tilde J_{\bf k}-\tilde J_{\bf Q})- \mu\Big] \tilde a_{\bf   k}^\dag \tilde a_{\bf k} ,\\
\label{H2}
H_2&=&\frac{1}{8}\sum_{\bf q} \Big[(5\tilde J_{\bf q}-2\tilde J_{\bf Q})\psi_{1}^{2}\tilde a_{\bf Q+q}^\dag \tilde a_{\bf Q-q}+\bf h.c.\Big]\nonumber\\
&&+\sum_{\bf k}(\tilde J_{\bf 0}-\tilde J_{\bf Q}+\tilde J_{\bf Q-k}-\tilde J_{\bf k})|\psi_{1}|^2 \tilde a_{\bf   k}^\dag \tilde a_{\bf k},
\end{eqnarray}
where, we remind, $\tilde J_{\bf k}$, where $\tilde{J}_{\bf k} = 2(J \cos[k_x] + 2 J'\cos[\frac{k_x}{2}] \cos[\frac{\sqrt{3} k_y}{2}])$. $\tilde{J}_{\bf k}$  has minimum $\tilde J_{\bf Q}$  at
$ {\bf Q}=(Q_{\rm i},0)$, and $ Q_{\rm i}=2\cos^{-1}[-J'/2J]$.  At small $\delta J \sim \delta J_c$, ${\bf Q}$ by $\bf{Q}\approx \rm  (4\pi/3-\Delta Q,0)$,
where $ \Delta Q= 4\pi/3 - Q_{\rm i} = 2\delta J/ {\sqrt{3}}$.

Our goal is to obtain the renormalization of the excitation spectrum $\omega_k$ to second order in the condensate, i.e., to order $\psi^2$.
The first term in $H_2$ is irrelevant for this purpose as it describes excitations with momentum transfer $2{\bf Q}$,which can only contribute to
   $\omega_k$ at second order in perturbation theory, but such term will be of order $\psi^4$.
      The remaining term in $H_2$ is quadratic in non-condensed bosons and directly contribute to spin-wave spectrum to second order in $\psi$

 We will be interested in magnon excitations for ${\bf k}$ near  $-{\bf Q}=-(Q_{\rm i},0)$. Accordingly, we set  ${\bf k}=-{\bf Q}+{\bf p}$ and treat ${\bf p}$ as small momentum. 
 Restricting with small ${\bf p}$ and using the approximate form of ${\bf Q}$,  we re-write  Eqs.\eqref{H1} and \eqref{H2} as
\begin{eqnarray}
{\cal H}_{\rm anis}=\sum_{\bf p}\Big[ \frac{3}{4}SJ (p_{x}^{2}+p_{y}^{2}) + J |\psi_{1}|^2\Big(\frac{h_{\rm sat}}{SJ}+\frac{9}{2}p_{x}\Delta Q+\frac{27}{4}(\Delta Q)^2\Big)-\mu \Big]
\tilde a_{\bf   -Q+p}^\dag \tilde a_{\bf -Q+p} ,
\end{eqnarray}
where $h_{\rm sat}=S(\tilde J_{\bf 0}-\tilde J_{\bf Q})=S\Gamma_{1}^{(0)}$. Completing the square and rearranging, and setting ${\bf k}=-{\bf Q}+{\bf p}$ again, we obtain
\begin{equation}
\label{eff H}
{\cal H}_{\rm anis}=\sum_{\bf k}S\omega_{\bf k}^{(1)}\tilde a_{\bf  k}^\dag \tilde a_{\bf k},
\end{equation}
where
\begin{eqnarray}
\omega_{\bf k}^{(1)}&=&\frac{3}{4}J\Big[(k_{x}+{\bar Q}_{\rm i})^{2}+k_{y}^{2}+\varepsilon_{\rm min}\Big],\\
\label{energy}
\varepsilon_{\rm min}&=&9\frac{|\psi_{1}|^2}{S}(1-\frac{|\psi_{1}|^2}{S})(\Delta Q)^2 + \frac{4}{3}\frac{1}{SJ}(\frac{|\psi_{1}|^2}{S}h_{\rm sat}-\mu).
\end{eqnarray}
Here  ${\bar Q}_{\rm i} =4\pi/3 - \Delta Q +3|\psi_{1}|^2\Delta Q/S$, and the minimum of $\omega_{\bf k}^{(1)}$ is at $(-{\bar Q}_{\rm i}, 0)$.

In the classical approximation (leading order in $1/S$), the condensate density is $|\psi_{1}|^2/S = \mu/(S \Gamma_{1}^{(0)}) = \mu/h_{\rm sat}$,
and we obtain
\begin{equation}
\varepsilon_{\rm min, class}=\frac{12\mu}{h_{\rm sat}J^{2}}\frac{h}{h_{\rm sat}} (\delta J)^2.
\label{eq:emin0}
\end{equation}
To this order, the second term in \eqref{energy} nullifies exactly.
 To the same accuracy, ${\bar Q}_{\rm i} = Q_{\rm i} + (4 \pi/3 -Q_{\rm i}) (3\mu/h_{\rm sat}) + O(1/S)$.

\subsection{Quantum corrections}

Since at $\mu=0$ the critical value of $\delta J_c \sim 1/\sqrt{S}$, we recognize that in fact $\varepsilon_{\rm min, class} \sim 1/S$ in the relevant range
of $\delta J$,
  where the transition between the cone and the coplanar state takes place. This means that  Eq.\eqref{eq:emin0} is not complete --
one needs to add to it quantum $1/S$ contributions. These come from several sources as we now describe.

The first
 quantum
 correction comes from the fact that
  the relation between the condensate wave function $\psi_1$ and
  $\Gamma_1$:
  \begin{equation}
\frac{|\psi_{1}|^2}{S}=\frac{\mu }{S\Gamma_{1}}
\end{equation}
contains $1/S$ terms because
   $\Gamma_1 = \Gamma_1^{(0)} + \Gamma_1^{(1)}$, where
$\Gamma_1^{(0)} = h_{\rm sat}/S = \tilde J_{\bf 0}-\tilde J_{\bf Q}\sim J$ represents classical ($S=\infty$) contribution already accounted for in deriving \eqref{eq:emin0},
while $\Gamma_1^{(1)} = \Gamma_{1a}^{(1)} + \Gamma_{1b}^{(1)} \sim J/S$ represents the leading $1/S$ correction to it. The term with subindex $a$ describes
contribution from normal ordering, $-J_{\bf Q}/(8S)$ in \eqref{eq:delta gam3}, while
 the one
  with subindex $b$ describes the contribution   from quantum fluctuations, Eq.\eqref{eq:gamma12-quantum}.

Hence, in the cone state,
\begin{equation}
\frac{|\psi_{1}|^2}{S}=\frac{\mu }{S\Gamma_{1}}=   \frac{\mu }{S(\Gamma_1^{(0)} + \Gamma_1^{(1)})} = \frac{\mu }{h_{\rm sat}}(1 - \frac{\Gamma_1^{(1)}}{\Gamma_1^{(0)}})
\end{equation}
contains quantum correction, $\sim \Gamma_1^{(1)}$. Substituting the full form of $\psi$ into Eq. (\ref{energy}) and collecting $1/S$ terms we
obtain first $1/S$ correction $\Delta \varepsilon_{\rm min,1}$
\begin{equation}
\Delta \varepsilon_{\rm min,1}=
-\frac{4}{3}\frac{\mu}{h_{\rm sat}J}\Gamma_{1}^{(1)} + O(1/S^2),
\label{eq:emin1}
\end{equation}

The two other quantum corrections are associated with $\Gamma_2$ processes.
One $\Gamma_2$  correction comes from
 Eq.\eqref{eq:delta H4}, which, we remind, emerges when we normal order bosonic operators in the Holstein-Primakoff transformation. 
 It is easiest to obtain this contribution  via a real-space representation
\begin{equation}
a_{\bf r}=\psi_1 e^{i {\bf Q} \cdot {\bf r}}+\tilde a_{\bf r},
\end{equation}
where, as before, $\tilde a_{\bf r}$ describes non-condensate magnons. Substituting this into \eqref{eq:delta H4} we obtain
\begin{equation}
\delta{\cal H}^{\rm(4)}=-\frac{|\psi_{1}|^2}{8S}\sum_{\bf k}(\tilde J_{\bf k}+\tilde J_{\bf Q})\tilde a_{\bf k}^\dagger\tilde a_{\bf k}\approx  -\frac{|\psi_{1}|^2}{4S}\sum_{\bf p}\tilde J_{\bf Q} \tilde a_{\bf   -Q+p}^\dag \tilde a_{\bf -Q+p},
\end{equation}
for ${\bf k} \approx - {\bf Q}$. Adding this to \eqref{energy} we obtain a $\Gamma_{2a}^{(1)}$ correction to  $\varepsilon_{\rm min}$,
\begin{equation}
\label{dmin2}
\Delta\varepsilon_{\rm min,2}= \frac{4}{3 J}\frac{\mu}{h_{\rm sat}J} (- \frac{\tilde J_{\bf Q}}{4S}) =\frac{4}{3}\frac{\mu}{h_{\rm sat}J}\Gamma_{2a}^{(1)}.
\end{equation}
Observe that, because we already have $1/S$ in the prefactor, we can neglect  the difference between $\tilde J_{\bf Q}$ and $J_{\bf Q}$.

The
 third quantum correction (also associated with $\Gamma_2$) comes from terms cubic in non-condensate magnons ${\tilde a}_{\bf k}$ taken to second order in perturbation theory.
 The cubic terms are generated from \eqref{eq:H^4} via the substitution \eqref{eq:cone-a}.
Such terms are necessarily linear in $\psi_1$:
\begin{equation}
\label{H3}
H_3=\frac{1}{\sqrt N}\sum_{\bf k,q} V_{\bf q}({\bf k,Q})\Big(\psi_1 \tilde a_{\bf   Q-q}^\dag \tilde a_{\bf k+q}^\dag \tilde a_{\bf k}+\bf h.c.\Big).
\end{equation}
 A second-order in perturbation theory in \eqref{H3} produces a  $1/S$ correction to the dispersion of ${\tilde a}_{\bf k}$
magnons with ${\bf k} \approx -{\bf Q}$ in the form
\begin{eqnarray}
\Delta\varepsilon_{\rm min,3}&=&-\frac{4}{3}\frac{1}{JS}\sum_{\bf q,q'} V_{\bf q}({\bf-Q,Q})V_{\bf q'}({\bf-Q,Q})|\psi_{1}|^2\langle \tilde a_{\bf   Q-q}^\dag \tilde a_{\bf -Q+q}^\dag \tilde a_{\bf   Q-q'}\tilde a_{\bf -Q+q'}\rangle_0,\nonumber\\
\label{dmin3}
&=& \frac{4}{3}\frac{|\psi_{1}|^2}{SJ} \Gamma_{2b}^{(1)}\approx\frac{4}{3}\frac{\mu}{h_{\rm sat}J} \Gamma_{2b}^{(1)}.
\end{eqnarray}
Adding Eqs. \eqref{eq:emin1}, \eqref{dmin2}, and \eqref{dmin3} to the classical result for $\varepsilon_{\rm min}$,  
 we obtain the  final expression for the minimal energy $\varepsilon_{\rm min}$
of the magnons at ${\bf k} \approx -{\bf Q}$:
\begin{equation}
\varepsilon_{\rm min,tot}=\frac{12\mu}{h_{\rm sat}J^2}\Big[\frac{h}{h_{\rm sat}}(\delta J)^2 + \frac{\Gamma_2^{(1)} - {\Gamma_1^{(1)}}}{9}\Big]
=\frac{12\mu}{h_{\rm sat}J^2}\Big[\frac{h}{h_{\rm sat}}(\delta J)^2-(\delta J_c)^2\Big]
\approx \frac{12\mu}{h_{\rm sat}J^2}\Big[(\delta J)^2-(\delta J_c)^2(1+\frac{\mu}{h_{\rm sat}})\Big].
\end{equation}
Observe that $\Gamma_{2}^{(1)}-\Gamma_{1}^{(1)} = \Delta \Gamma^{(1)} = -1.6J/S$ and $\delta J_c = \sqrt{1.6 J^2/(9 S)} \approx 0.42 J/\sqrt{S}$,
see Eq.\eqref{eq:d-gamma} and description below it in the main text.

At $\mu = +0$, magnon energy of ${\bf k} = - {\bf Q}_{\rm i}$ vanishes at $\delta J = \delta J_c$, as expected.
However, at a finite $\mu$, the instability occurs at  $\delta J _h = h_{\rm sat} \delta J_{c}/h > \delta J_{c}$
and the mode that condenses carries momentum  ${\bf k} = (-{\bar Q}_{\rm i},0) \neq - {\bf Q}_{\rm i}$ different from $- {\bf Q}_{\rm i}$.
This gives rise to the development of the second condensate
with momentum $(-{\bar Q}_{\rm i},0)$. The resulting state is the double cone phase described in the main text.

\end{widetext}

\end{document}